\documentstyle[12pt]{article}

\newcommand{\sect}[1]{\setcounter{equation}{0}\section{#1}}

\newcommand{\bea}{\begin{eqnarray}}

\newcommand{\eea}{\end{eqnarray}}

\newcommand{\be}{\begin{equation}}

\newcommand{\ee}{\end{equation}}

\newcommand{\vs}[1]{\vspace{#1 mm}}

\renewcommand{\a}{\alpha}

\renewcommand{\b}{\beta}

\renewcommand{\d}{\delta}

\newcommand{\dsl}{\pa \kern-0.5em /}

\newcommand{\pa}{\partial}

\renewcommand{\t}{\theta}

\newcommand{\nn}{\nonumber\\}

\begin{document}

\topmargin 0pt

\oddsidemargin 0mm



\begin{flushright}

USTC-ICTS-05-4\\


hep-th/0503007\\


\end{flushright}

\vspace{2mm}

\begin{center}

{\Large \bf

Non-SUSY $p$-branes delocalized in two directions,\\
tachyon condensation and T-duality}

\vs{6}

{\large J. X. Lu$^a$\footnote{E-mail: jxlu@ustc.edu.cn}
 and Shibaji Roy$^b$\footnote{E-mail: roy@theory.saha.ernet.in}}

 \vspace{4mm}

{\em

 $^a$ Interdisciplinary Center for Theoretical Study\\

 University of Science and Technology of China, Hefei, Anhui
 230026, China\\
 and \\
 Center for Mathematics and Theoretical Physics\\
 Shanghai Institute for Advanced Study\\
 University of Science and Technology of China, Shanghai, China\\

and\\

Interdisciplinary Center of Theoretical Studies\\

Chinese Academy of Sciences, Beijing 100080, China\\




\vs{4}

 $^b$ Saha Institute of Nuclear Physics,

 1/AF Bidhannagar, Calcutta-700 064, India}

\end{center}

\vs{4}

\centerline{{\bf{Abstract}}}

\vs{4}

\begin{small}
We here generalize our previous construction [hep-th/0409019] of
non-supersymmetric $p$-branes delocalized in one transverse
spatial direction to two transverse spatial directions in
supergravities in arbitrary dimensions ($d$). These solutions are
characterized by five parameters. We show how these solutions in
$d=10$ interpolate between D$(p+2)$-anti-D$(p+2)$ brane system,
non-BPS D$(p+1)$-branes (delocalized in one direction) and BPS
D$p$-branes by adjusting and scaling the parameters in suitable
ways. This picture is very similar to the descent relations
obtained by Sen in the open string effective description of
non-BPS D$(p + 1)$-brane and BPS D$p$-brane as the respective
tachyonic kink and vortex solutions on the D$(p+2)$-anti-D$(p+2)$
brane system (with some differences). We compare this process
with the T-duality transformation which also has the effect of
increasing (or decreasing) the dimensionality of the branes by
one.
\end{small}

\newpage

\section{Introduction}

In \cite{luroyone} we constructed delocalized, non-supersymmetric
$p$-brane solutions of maximal supergravities in arbitrary
dimensions ($d$). In $d=10$, these solutions in type IIA (IIB)
theory (for $p$ = even (odd)) were interpreted as interpolating
solutions between non-BPS D$(p+1)$-branes and a codimension one
BPS D$p$-branes when the transverse delocalized direction was
spatial. To certain extent, this picture of supergravity captures
how the tachyonic kink solution on the non-BPS D$(p+1)$-brane
\cite{asone} from the open string effective description emerges.
The picture holds even for the temporally delocalized Euclidean
$p$-brane solutions which was obtained from the spatially
delocalized solution by a Wick rotation \cite{astwo,oy,luroyone}.
The emergence of this picture similar to the open string tachyon
condensation might seem surprising as there is no explicit
appearance of the tachyon field in the supergravity description.
We, however, take it seriously and by going a step further
construct the non-supersymmetric $p$-brane solutions delocalized
in two transverse spatial directions in this paper. One of the
main motivations for this construction was to see how we can
understand the descent relations obtained by Sen
\cite{asthree,asone} in the open string effective description of
BPS and non-BPS D-branes as tachyonic soliton solutions on
brane-antibrane pair of higher dimension from the supergravity
point of view. In fact, we will show how these solutions can be
regarded as interpolating solutions between D$(p+2)$-${\bar {\rm
D}}(p+2)$ brane systems, non-BPS D$(p+1)$-branes (delocalized in
one spatial transverse direction) and localized BPS D$p$-branes,
similar to the descent relations of Sen for the tachyonic kink
(vortex) solutions on the D-${\bar{\rm D}}$-brane systems
\cite{asone}. We will also study the T-duality properties of these
solutions.

The descent relation D$(p+2)$-${\bar {\rm D}}(p+2)$ $\to$ non-BPS
D$(p+1)$ $\to$ BPS D$p$ obtained in the course of finding the
tachyonic kink solution on the previous system involves branes on
the two sides whose dimensionalities differ by two. In the absence
of explicit tachyon field, the most natural way to see this
picture emerging in the supergravity solution is to consider
non-supersymmetric $p$-branes delocalized in two transverse
spatial directions. This is what we study in this paper. For the
case of BPS D$p$-branes the delocalized solutions \cite{lps,km}
are obtained by first periodically placing an infinite array of
branes along the transverse directions (it can be done in steps
when more than one directions are to be delocalized) and then
taking the continuum limit\footnote{Usually this produces
isometries along the transverse directions of the brane and then
the application of T-duality along those directions gives the
localized higher dimensional branes.}. This procedure assumes the
`no-force' condition among the BPS branes. However, for the
non-supersymmetric branes, since they interact with each other it
is not clear how this procedure will work. In \cite{luroyone}, we
obtained the delocalized, non-supersymmetric $p$-branes by
explicitly solving the equations of motion with an appropriate
metric ansatz. This gives entirely new solutions whose relation
with the localized non-supersymmetric $p$-branes is not at all
obvious unlike the BPS case.

In this paper we generalize the delocalized solutions from one
transverse spatial direction to two transverse directions. The
generalization is non-trivial and one needs to solve the equations
of motion again to obtain these solutions. We first write the
solutions in $d$-dimensions which are characterized by five
independent parameters. We show in $d=10$ that the solutions can
be made localized $(p+2)$-branes in two ways when the parameters
satisfy certain conditions. In the first case, when there is no
T-duality involved, the resulting solutions can be interpreted as
D$(p+2)$-${\bar {\rm D}}(p+2)$ with zero net charge and in the
second case when we apply T-duality (studied in section 4) twice along 
the delocalized
directions the resulting solutions can be interpreted as charged
D$(p+2)$-${\bar {\rm D}}(p+2)$ solutions. While in the first case
the solutions will be characterized by two parameters, in the
second case they will be characterized by three parameters. Next
we will show that when the parameters satisfy some other
conditions, we can convert these solutions to delocalized
$(p+1)$-branes (note that unlike in the previous case we can not
make them completely localized) again in two ways. When no
T-duality is involved the solutions can be interpreted as non-BPS
D$(p+1)$-branes (delocalized in one direction) characterized by
three parameters and when we take T-duality in one of the
delocalized transverse directions, the solutions can be
interpreted as charged D$(p+1)$-${\bar {\rm D}}(p+1)$ (delocalized
in one direction) solutions characterized by four parameters\footnote{
We do not exactly study this case as mentioned in footnote 8.
Instead we apply T-duality once to the non-susy $p$-brane solutions
delocalized in one transverse direction \cite{luroyone} and
obtain charged D$(p+1)$-${\bar{\rm D}}(p+1)$ brane system which
is fully localized (see section 4).}. Note
that in the latter case since we take T-duality once the theory
changes from type IIA (IIB) to IIB (IIA). Finally, we show that by
appropriately scaling the parameters we can convert the solutions
to localized BPS D$p$-branes. So, when no T-duality is involved we
interpret our solutions as interpolating solutions between
D$(p+2)$-${\bar {\rm D}}(p+2)$ brane system (with net charge
zero), non-BPS D$(p+1)$-branes (delocalized in one direction) and
BPS D$p$-branes very similar to the descent relations advocated by
Sen \cite{asone} in the open string effective description of
non-BPS D$(p + 1)$-brane and BPS D$p$-brane as the respective
tachyonic kink and vortex solutions on the D$(p+2)$-anti-D$(p+2)$
brane system.

As we mentioned the process in obtaining the non-BPS D-brane (BPS
D-brane) as the tachyonic kink (and vortex) solutions on the
brane-antibrane systems has some similarities with the T-duality
transformation. For example, for BPS D-branes, the following
transition, D$(p+2)$ $\to$ D$(p+1)$ $\to$ D$p$ can be obtained by
T-duality along the brane directions at each step. However, the
crucial difference is that while for T-duality the theory changes
from type IIA (IIB) to IIB (IIA) at each step, the above process
does not change the theory. We will perform T-duality on the
delocalized solutions in a separate section for comparison.

This paper is organized as follows. In section 2, we write down the
non-supersymmetric $p$-brane solutions delocalized in two transverse
spatial directions in arbitrary dimensions. In section 3, we show how
in $d=10$, these solutions nicely interpolate between
D$(p+2)$-${\bar {\rm D}}(p+2)$ brane system, non-BPS D$(p+1)$-branes
(delocalized in one transverse direction) and the localized BPS D$p$-branes
similar to the descent relation of Sen. In section 4, we study the
T-duality transformation of the delocalized solutions for comparison. Our
conclusion is presented in section 5.

\sect{The delocalized solutions}

In this section we give the non-supersymmetric $p$-brane solutions
delocalized in two transverse spatial directions of maximal
supergravities in arbitrary space-time dimensions $d$. This is a
generalization of the delocalized solutions given in
\cite{luroyone} from one transverse direction to two. The
generalization is non-trivial and to obtain them one has to solve
the equations of motion following from the effective action with
an appropriate metric ansatz. The $d$-dimensional supergravity
action we consider has the form, \be S = \int d^dx \sqrt{-g}
\left[R - \frac{1}{2}
\partial_\mu \phi \partial^\mu \phi - \frac{1}{2\cdot q!}
e^{a\phi} F_{[q]}^2\right] \ee where $g_{\mu\nu}$, with $\mu,\,\nu
= 0,1,\ldots,d-1$, is the metric and $g={\rm det}(g_{\mu\nu})$,
$R$ is the scalar curvature, $\phi$ is the dilaton, $F_{[q]}$ is
the field strength of a $(q-1)=(d-p-3)$-form gauge field and $a$
is the dilaton coupling. The action (2.1) represents the bosonic
sector of the low energy effective action of string/M theory
dimensionally reduced to $d$-dimensions. Now in order to obtain
the delocalized solutions in two transverse directions, we have to
solve the equations of motion from (2.1) with the following ansatz
for the metric and the $q$-form field strength, \bea ds^2 &=&
e^{2A(r)}\left(dr^2 + r^2 d\Omega_{d-p-4}^2\right) + e^{2B(r)}
\left( -dt^2 + \sum_{i=1}^p dx_i^2 \right) + e^{2C(r)} dx_{p+1}^2
+ e^{2D(r)} dx_{p+2}^2\nn F_{[q]} &=& b\,\, {\rm
Vol}(\Omega_{d-p-4}) \wedge dx_{p+1} \wedge dx_{p+2} \eea In the
above $r = (x_{p+3}^2 + \cdots + x_{d-1}^2)^{1/2}$,
$d\Omega_{d-p-4}^2$ is the line element of a unit
$(d-p-4)$-dimensional sphere, Vol($\Omega_{d-p-4}$) is its
volume-form and $b$ is the magnetic charge parameter. The
solutions (2.2) represent magnetically charged $p$-brane solutions
delocalized in transverse $x_{p+1}$ and $x_{p+2}$ directions. The
equations of motion will be solved with the following gauge
condition, \be (p+1) B(r) + (q-3) A(r) + C(r) + D(r) = \ln G(r)
\ee Note that as $G(r) \to 1$, the above condition reduces to the
extremality or the supersymmetry condition \cite{dkl}. As
mentioned in \cite{luroytwo}, the consistency of the equations of
motion dictates that the non-extremality function $G(r)$ can take
three different forms and we will need only one of them for our
purpose which is, \be G(r) = 1-\frac{\omega^{2(q-3)}}{r^{2(q-3)}}
= \left(1+\frac{\omega^{q-3}}
{r^{q-3}}\right)\left(1-\frac{\omega^{q-3}} {r^{q-3}}\right) =
H(r) \tilde{H}(r) \ee By solving the equations of motion following
from (2.1) with the ansatz (2.2) we obtain, \bea A(r) &=&
-\frac{p+1}{q-1}B(r) - \ln\left(\frac{H}{\tilde H}\right)^{\frac{
\delta_2+\delta_3}{q-3}} + \ln\left(H {\tilde
H}\right)^{\frac{1}{q-3}}\nn C(r) &=& -\frac{p+1}{q-1}B(r) +
\delta_2 \ln \left(\frac{H}{\tilde H}\right) \nn D(r) &=&
-\frac{p+1}{q-1}B(r) + \delta_3 \ln \left(\frac{H}{\tilde
H}\right)\nn \phi(r) &=& \frac{a(d-2)}{q-1}B(r) + \delta_1 \ln
\left(\frac{H}{\tilde H} \right) \nn B(r) &=& - \frac{2}{\chi} \ln
F(r) \eea where $\chi = 2(p+1) + a^2(d-2)/(q-1)$ and the function
$F(r)$ is defined as \be F(r) = \cosh^2\theta
\left(\frac{H}{\tilde H}\right)^\alpha - \sinh^2\theta
\left(\frac{\tilde H}{H}\right)^\beta \ee In eqs.(2.4), (2.5),
$\omega$, $\delta_1$, $\delta_2$, $\delta_3$ are real integration
constants and in (2.6), $\theta$, $\alpha$, $\beta$ are some real
parameters. However, not all of them are independent. The
parameter relations are given in the following, \bea
&&\alpha-\beta \,\,\,= \,\,\,a\delta_1\\
&&\frac{1}{2} \delta_1^2 + \frac{2\alpha(\alpha-a\delta_1)(d-2)}{\chi(q-1)} +
\frac{2 \delta_2\delta_3}{q-3} \,\,\,=\,\,\, (1-\delta_2^2 - \delta_3^2)
\frac{q-2}{q-3}\\
&&b\,\,\, =\,\,\, \sqrt{\frac{4(d-2)}{(q-1)\chi}} (q-3)
\omega^{q-3} (\alpha+\beta) \sinh2\theta \eea So, from (2.7), we
can determine $\beta$ in terms of $\alpha$ and $\delta_1$. Also,
from (2.8), we can determine $\alpha$ in terms of $\d_1$, $\d_2$
and $\d_3$. Eq.(2.9) gives a relation between the charge parameter
$b$ with the other parameters. So, the solution (2.5), (2.6)
contains only five independent parameters, namely, $\omega$,
$\theta$, $\d_1$, $\d_2$ and $\d_3$. Writing explicitly the
non-susy $p$-brane solutions delocalized in two transverse
directions in $d$ space-time dimensions, we have, \bea ds^2 &=&
F^{\frac{4(p+1)}{(q-1)\chi}} (H{\tilde {H}})^{\frac{2}{q-3}}
\left(\frac{H}{\tilde
H}\right)^{-\frac{2(\d_2+\d_3)}{q-3}}\left(dr^2 + r^2
d\Omega_{d-p-4}^2\right) + F^{-\frac{4}{\chi}}\left(-dt^2 +
\sum_{i=1}^p dx_i^2 \right)\nn & & \qquad\qquad +
F^{\frac{4(p+1)}{(q-1)\chi}} \left(\frac{H}{\tilde
H}\right)^{2\d_2} dx_{p+1}^2 + F^{\frac{4(p+1)}{(q-1)\chi}}
\left(\frac{H}{\tilde H}\right)^{2\d_3} dx_{p+2}^2\nn e^{2\phi}
&=& F^{-\frac{4a(d-2)}{(q-1)\chi}} \left(\frac{H}{\tilde
{H}}\right)^{2\delta_1},\qquad F_{[q]}\,\,\, =\,\,\, b\,\,{\rm
Vol}(\Omega_{d-p-4})\wedge dx_{p+1} \wedge dx_{p+2} \eea These are
magnetically charged, non-supersymmetric, delocalized $p$-brane
solutions and the corresponding electrically charged solutions can
be obtained by replacing $g_{\mu\nu} \to g_{\mu\nu}$, $\phi \to
-\phi$, $F \to e^{-a\phi} \ast F$. We note from the form of
${\tilde H}(r)$ given in (2.4) that the solutions above have
potential singularities at $r=\omega$ and for avoiding this
complication, we limit our discussion to the well-defined region
of $r>\omega$. Also, as $r \to \infty$, $H$, $\tilde H$, $F$ $\to
1$ and so the solutions are asymptotically flat. We like to point
out that if we send $H$, $\tilde H$ $\to 1$, such that the
function $F(r)$ reduces to the usual harmonic function of a BPS
$p$-brane then the above solutions indeed reduce to the BPS
$p$-branes delocalized in two transverse directions. The
delocalized BPS $p$-brane solutions can be made localized by the
usual procedure of replacing the extended source by a delta
function source \cite{bmm}. However, this procedure does not work
for the delocalized non-supersymmetric brane solutions given in
(2.10). In the following sections, we will see how a fully
localized $(p+2)$-brane solutions, a delocalized $(p+1)$-brane
solutions as well as a fully localized $p$-brane solutions can be
obtained from the delocalized solutions (2.10) with or without the
application of T-duality. In this process we will interpret the
above solutions as the interpolating solutions between
D$(p+2)$-${\bar{\rm D} }(p+2)$-brane systems, non-BPS
D$(p+1)$-branes delocalized in one transverse direction and BPS
D$p$-branes in $d=10$.

\sect{Delocalized solutions as interpolating solutions}

In this section we will show how the solutions given in (2.10) can be regarded
as interpolating solutions between D$(p+2)$-${\bar{\rm D}}(p+2)$-brane systems,
non-BPS D$(p+1)$-brane delocalized in one direction and BPS D$p$-branes similar
to the descent relation advocated by Sen. For this purpose we restrict our
solutions in $d=10$ and rewrite (2.10) as follows,
\bea
ds^2 &=& F^{\frac{p+1}{8}} (H{\tilde {H}})^{\frac{2}{5-p}}
\left(\frac{H}{\tilde H}\right)^{-\frac{2(\d_2+\d_3)}{5-p}}\left(dr^2 + r^2
d\Omega_{6-p}^2\right)
+ F^{-\frac{7-p}{8}}\left(-dt^2 + \sum_{i=1}^p dx_i^2
\right)\nn
& & \qquad\qquad
+ F^{\frac{p+1}{8}}
\left(\frac{H}{\tilde H}\right)^{2\d_2} dx_{p+1}^2
+ F^{\frac{p+1}{8}}
\left(\frac{H}{\tilde H}\right)^{2\d_3} dx_{p+2}^2\nn
e^{2\phi} &=& F^{-a}
\left(\frac{H}{\tilde {H}}\right)^{2\delta_1},\qquad
F_{[8-p]}\,\,\, =\,\,\, b\,\,{\rm Vol}(\Omega_{6-p})\wedge dx_{p+1} \wedge
dx_{p+2}
\eea
where $F(r)$ is as given in (2.6) with $H=1+\omega^{5-p}/r^{5-p}$, $\tilde H
= 1- \omega^{5-p}/r^{5-p}$, $a=(p-3)/2$ for D$p$-branes and $a=(3-p)/2$
for NSNS branes. The parameter relations (2.7) -- (2.9) take the forms,
\bea
&&\alpha-\beta\,\,\, = \,\,\,a\delta_1\\
&&\frac{1}{2} \delta_1^2 + \frac{1}{2}\alpha(\alpha-a\delta_1) +
\frac{2 \delta_2\delta_3}{5-p}\,\,\, =\,\,\, (1-\delta_2^2 - \delta_3^2)
\frac{6-p}{5-p}\\
&&b \,\,\,=\,\,\, (5-p) \omega^{5-p} (\alpha+\beta) \sinh2\theta
\eea Using (3.2) and (3.3) we can express the parameters $\a$,
$\b$ in terms of three unknown parameters $\d_1$, $\d_2$ and
$\d_3$ as, \bea \a &=& \pm\sqrt{2\left(1-\d_2^2-\d_3^2 -
\frac{2\d_2\d_3}{6-p}\right) \frac{6-p}{5-p} -
\frac{(p+1)(7-p)}{16}\d_1^2} + \frac{a\d_1}{2}\nn \beta &=&
\pm\sqrt{2\left(1-\d_2^2-\d_3^2 - \frac{2\d_2\d_3}{6-p}\right)
\frac{6-p}{5-p} - \frac{(p+1)(7-p)}{16}\d_1^2} - \frac{a\d_1}{2}
\eea So, the solutions (3.1) are dependent on five parameters
$\omega$, $\theta$, $\d_1$, $\d_2$ and $\d_3$. Since the solutions
here are non-supersymmetric, the parameters are presumably be
related to the mass, charge, the tachyon vev $\langle T \rangle$
and the vev of its derivatives along the delocalized directions
$x_{p+1}$, $x_{p+2}$ i.e. $\langle \partial_1 T\rangle$ and
$\langle \partial_2 T\rangle$ of the brane systems. However, the
exact relationships between them are not clear to us. Now from
(3.5) we find that the parameters $\d_1$, $\d_2$, $\d_3$ must
satisfy the following bounds, \bea && \d_2^2 + \d_3^2 +
\frac{2\d_2 \d_3}{6-p} \,\,\, \leq \,\,\, 1\nn && |\d_1| \,\,\,
\leq \,\,\, 4 \sqrt{\frac{2(6-p)}{(5-p)(p+1)(7-p)}}
\sqrt{\left(1-\d_2^2 - \d_3^2 - \frac{2\d_2\d_3}{6-p}\right)} \eea
Once we know the form of the metric given in (3.1), we can compute
the energy-momentum (e-m) tensor associated with the brane from
the linearized form of the Einstein equation given by, \be
\nabla^2 \left(h_{\mu\nu} - \frac{1}{2} \eta_{\mu\nu} h\right) =
-2\kappa_0^2 T_{\mu\nu} \delta^{(7-p)} (r) \ee where we have
expanded the metric around asymptotically flat space as
$g_{\mu\nu} = \eta_{\mu\nu} + h_{\mu\nu}$ and used the harmonic
gauge $\partial_\lambda h_\mu^\lambda - \frac{1}{2} \partial_\mu h
= 0$ with $h=\eta^{\mu\nu} h_{\mu\nu}$. Also in (3.7) $2\kappa_0^2
= 16 \pi G_{10}$, $G_{10}$, being the ten dimensional Newton's
constant. So, from the metric in (3.1) and (3.7) we obtain the
various components of e-m tensor as, \bea T_{00} &=&
\frac{\Omega_{6-p}}{2\kappa_0^2} (5-p) \omega^{5-p} \left[(\a +
\beta)\cosh2\t + (\a - \beta) - \frac{4(\d_2+\d_3)}{5-p}\right]\nn
T_{ij} &=& -\frac{\Omega_{6-p}}{2\kappa_0^2} (5-p) \omega^{5-p}
\left[(\a + \beta)\cosh2\t + (\a - \beta) -
\frac{4(\d_2+\d_3)}{5-p}\right] \delta_{ij}\nn T_{p+1,p+1} &=&
\frac{\Omega_{6-p}}{2\kappa_0^2} (5-p) \omega^{5-p}
\left[\frac{4\d_2(6-p) + 4\d_3}{5-p}\right]\nn T_{p+2,p+2} &=&
\frac{\Omega_{6-p}}{2\kappa_0^2} (5-p) \omega^{5-p}
\left[\frac{4\d_3(6-p) + 4\d_2}{5-p}\right]\nn T_{mn} &=& 0 \eea
In the above $i,j=1,\ldots,p$ and $m,n=p+3,\ldots,9$. $\Omega_n =
2\pi^{(n+1)/2}/\Gamma((n+1)/2)$ is the volume of the
$n$-dimensional unit sphere. $T_{00}$ in (3.8) is the mass per
unit brane volume. Since it has the dimensionality of the mass per
unit $(p+2)$-dimensional brane volume, we find that the energy is
spread also along the delocalized directions $x_{p+1}$ and
$x_{p+2}$ as expected\footnote{Note that the e-m tensors obtained
from boundary CFT or from string field theory as given in
eq.(3.51) of ref.\cite{asfour} differ from those given above in
the sense that the former involves a source function $f(x)$ for
$T_{00}$ and $T_{ij}$ whereas for us they are constant. It will be
interesting to find supergravity solutions which will produce such
functional dependence on the components of the e-m tensor.}. The
brane is spread along $x_{p+1}$ and $x_{p+2}$ can also be noted
from the non-vanishing components of e-m tensor $T_{p+1,p+1}$ and
$T_{p+2,p+2}$. $T_{mn}=0$ implies  that the brane is localized
along $x_{p+3}, \ldots, x_{d-1}$ directions and they are the true
transverse directions.

Now let us look at the metrics in (3.1). They represent the
metrics for non-supersymmetric $p$-branes delocalized in two
directions $x_{p+1}$, $x_{p+2}$ which are the isometric
directions. For the similar case of BPS solutions they are usually
made fully localized along the transverse directions by
T-dualities in both $x_{p+1}$ and $x_{p+2}$ directions. Here we
will see that the solutions in (3.1) can be made fully localized
$(p+2)$-branes both with and without making T-duality
transformations along $x_{p+1}$ and $x_{p+2}$ directions. We will
consider T-duality in the next section. However, in this section
no T-duality will be employed. We note from the metrics in (3.1)
that they can be made localized $(p+2)$-branes if the coefficient
of the term $(-dt^2+\sum_{i=1}^p dx_i^2)$ match with both
$dx_{p+1}^2$ and $dx_{p+2}^2$ terms. This is possible only if \bea
\theta &=& b \,\,\, = \,\,\, 0\nn \d_2 &=& \d_3\,\,\, = \,\,\,
-\frac{\a}{2} \eea Since we have $b=0$, the solution is chargeless
and in that case we see from (2.6) that the function $F(r)$
simplifes to $F(r) = \left(\frac{H}{\tilde H}\right)^\a$. The form
of the e-m tensor in this case can be obtained from (3.8) as, \bea
T_{00} &=& \frac{\Omega_{6-p}}{2\kappa_0^2} (5-p) \omega^{5-p}
\left[\frac{2\a (7-p)}{5-p}\right]\nn T_{ij} &=&
-\frac{\Omega_{6-p}}{2\kappa_0^2} (5-p) \omega^{5-p}
\left[\frac{2\a(7-p)}{5-p}\right] \delta_{ij}\nn T_{mn} &=& 0 \eea
where $i,j=1,2,\ldots,(p+2)$ and $m,n=p+3,\ldots,9$. This is
exactly what we expect of localized $(p+2)$-branes. The solutions
(3.1) in this case take the form \bea ds^2 &=& (H{\tilde
{H}})^{\frac{2}{5-p}} \left(\frac{H}{\tilde
H}\right)^{\frac{p+1}{8}\a + \frac{2\a}{5-p}} \left(dr^2 + r^2
d\Omega_{6-p}^2\right) + \left(\frac{H}{\tilde
H}\right)^{-\frac{7-p}{8}\a} \left(-dt^2 + \sum_{i=1}^{p+2} dx_i^2
\right)\nn e^{2\phi} &=& \left(\frac{H}{\tilde {H}}\right)^{-a\a +
2\delta_1},\qquad F_{[8-p]}\,\,\, =\,\,\, 0 \eea and the
parameters $\a$ and $\d_1$ now satisfy \be \d_1^2 - a\a \d_1 +
2(\a^2-1)\frac{6-p}{5-p} = 0 \ee We identify the above solutions
as D$(p+2)$-${\bar{\rm D}}(p+2)$-brane systems with zero net
charge \cite{luroythree,berto,luroytwo}. It should be remarked
here that since $F_{[8-p]}=0$, (3.11) can also represent non-BPS
D$(p+2)$-branes \cite{bmo} in the T-dual theory of the theory
 we start with in (2.1). With respect to our theory with a R-R ($8 - p$)-form
 field strength, our above configuration should represent the chargeless
D$(p+2)$-${\bar{\rm D}}(p+2)$ brane system. We also remark that
this localization is possible from the delocalized solutions (3.1)
without taking T-duality because of the presence of the extra
parameters in the solutions. Since these extra parameters are not
present in BPS solutions, localization in that case is possible
only through T-duality.

Next we will try to make the $p$-brane solutions in (3.1) to
$(p+1)$-brane solutions by equating the coefficients of the term
$(-dt^2+\sum_{i=1}^p dx_i^2)$ and $dx_{p+1}^2$. We find that this
is possible if we put  \bea b &=& \theta
\,\,\,=\,\,\, 0\nn \d_2 &=& -\frac{\a}{2}, \qquad
\d_3\,\,\,=\,\,\, {\rm arbitrary} \eea Using (3.13) we find that
the e-m tensors in (3.8) take the forms, \bea T_{00} &=&
\frac{\Omega_{6-p}}{2\kappa_0^2} (5-p) \omega^{5-p}
\left[\frac{2\a (6-p) - 4\d_3}{5-p}\right]\nn T_{ij} &=&
-\frac{\Omega_{6-p}}{2\kappa_0^2} (5-p) \omega^{5-p}
\left[\frac{2\a(6-p) - 4\d_3}{5-p}\right] \delta_{ij}\nn
T_{p+2,p+2} &=& -\frac{\Omega_{6-p}}{2\kappa_0^2} (5-p)
\omega^{5-p} \left[\frac{2\a - 4\d_3 (6-p)}{5-p}\right]\nn T_{mn}
&=& 0 \eea where now $i,j=1,\ldots,p+1$. It is clear from (3.14)
that when the parameters are restricted as (3.13), we get
$(p+1)$-brane solutions delocalized in $x_{p+2}$-direction. The
solutions (3.1) then take the forms, \bea ds^2 &=& (H{\tilde
{H}})^{\frac{2}{5-p}} \left(\frac{H}{\tilde
H}\right)^{\frac{p+1}{8}\a + \frac{\a-2\d_3}{5-p}} \left(dr^2 +
r^2 d\Omega_{6-p}^2\right)\nn &+& \left(\frac{H}{\tilde
H}\right)^{-\frac{7-p}{8}\a} \left(-dt^2 + \sum_{i=1}^{p+1} dx_i^2
\right) + \left(\frac{H}{\tilde H}\right)^{\frac{p+1}{8}\a +
2\d_3} dx_{p+2}^2\nn e^{2\phi} &=& \left(\frac{H}{\tilde
{H}}\right)^{-\frac{p-3}{2}\a + 2\delta_1},\qquad F_{[8-p]}\,\,\,
=\,\,\, 0 \eea and the parameters are related as, \be \frac{1}{2}
\d_1^2 + \frac{1}{2} \a(\a-a\d_1) - \frac{\a\d_3}{5-p} =
\left(1-\frac{\a^2}{4}-\d_3^2\right) \frac{6-p}{5-p} \ee It can be
easily checked that (3.15), (3.16) represent non-BPS
D$(p+1)$-branes delocalized in $x_{p+2}$ direction as obtained in
ref.\cite{luroyone}. One may think that these solutions can be
made localized if $T_{p+2,p+2}=0$ or in other words if \be \a =
2\d_3(6-p) \ee However, from the form of the metric in (3.15) it
is clear that even in this case we do not get a localized non-BPS
D$(p+1)$-brane solutions. We get misled because the e-m tensors
encode only the linear properties of the metric and not the full
metric. In fact it is easy to see that when (3.17) is satisfied
the $\left(\frac{H}{\tilde H}\right)$ factor in both the terms
$(dr^2 + r^2 d\Omega_{6-p}^2)$ and $dx_{p+2}^2$ match, but there
is an additional $(H{\tilde H})^{2/(5-p)}$ factor in front of the
first term which does not contribute to the linear term or the e-m
tensor, but forbids the metric to take a localized non-BPS
D$(p+1)$-brane form. Furthermore, we point out that even if we
ignore the non-linear part of $(H{\tilde H})$ factor, the metrics
in (3.1) can not be regarded as localized non-BPS D$(p+1)$ branes
because the parameter relation (3.16) differs from that of a
localized non-BPS brane solutions \cite{luroytwo}.

Now we will see how the delocalized solutions in (3.1) reduce to
BPS $p$-branes. The necessary condition for this to happen is that
the e-m tensor in (3.8) must satisfy $T_{00}=-T_{ii}$, for
$i=1,\ldots,p$ and $T_{mm}=0$ for $m=p+1,\ldots,9$. From (3.8) we
find that this condition implies that either $\d_2,\d_3 \to 0$,
or, $\omega^{5-p} \to 0$, $\theta \to \infty$ such that
$\omega^{5-p} \cosh 2\theta = $ finite. Examining the metric in
(3.1) carefully, we find that we have the correct BPS limits if we
scale the parameters as \bea \omega^{5-p} &\to& \epsilon {\bar
\omega}^{5-p}\nn (\a + \b) \sinh\,2\theta &\to& \epsilon^{-1} \eea
where $\epsilon \to 0$ is a dimensionless parameter. Also from
(3.4), (2.6) we find that under the above scaling $b \to (5-p)
{\bar \omega}^{5-p}$, $F \to {\bar H} = 1 + {\bar
\omega}^{5-p}/r^{5-p}$ and $H,\,{\tilde H} \to 1$. Now since
$\d$'s are bounded given by (3.6), it can be easily checked that
the configurations (3.1) reduce to \bea ds^2 &=& {\bar
{H}}^{\frac{p+1}{8}}\left(dr^2 + r^2 d\Omega_{6-p}^2 + dx_{p+1}^2
+ dx_{p+2}^2\right) + {\bar {H}}^{-\frac{7-p}{8}}\left( -dt^2 +
\sum_{i=1}^p dx_i^2\right)\nn e^{2\phi} &=& {\bar {H}}^{-a},
\qquad F_{[8-p]}\,\,\,=\,\,\, b {\rm Vol}(\Omega_{6-p})\wedge
dx_{p+1} \wedge dx_{p+2} \eea This is the BPS D$p$-brane solutions
delocalized in $x_{p+1}$ and $x_{p+2}$ directions. The components
of e-m tensor in (3.8) take the forms, $T_{00} = - T_{ii} =
\frac{\Omega_{6-p}}{2\kappa_0^2} (5-p) {\bar \omega}^{5-p}$ for
$i=1,\ldots,p$ and $T_{p+1,p+1} = 0$, $T_{p+2,p+2}=0$, $T_{mm}=0$.
Although the BPS configuration we get in (3.19) is delocalized,
this delocalization is trivial as opposed to the delocalized
solutions we got for non-BPS D$(p+1)$-branes in (3.15). This is
because we can localize the above solutions by replacing the
membrane-like source along $x_{p+1}$, $x_{p+2}$ by a point source
or delta function source without any cost of energy (true for BPS
branes). In calculating the e-m tensor we replace the Poisson's
equation of the harmonic function ${\bar H}(r)$ as \cite{bmm} \be
\nabla^2 {\bar H} = -\Omega_{6-p} (5-p) {\bar \omega}^{5-p}
\delta^{(7-p)}(r) \quad \Rightarrow \quad \nabla^2 {\bar H} =
-\Omega_{8-p} (7-p) {\bar \omega}^{7-p} \delta^{(9-p)}(r) \ee The
harmonic function now takes the form ${\bar H}(r) = 1+ {\bar
\omega}^{7-p} /r^{7-p}$, where $r$ includes $x_{p+1}$ and
$x_{p+2}$. The components of e-m tensors will be given as $T_{00}
= -T_{ii} = \frac{\Omega_{8-p}}{2\kappa_0^2} (7-p) {\bar
\omega}^{7-p}$, for, $i=1,\ldots, p$ and $T_{mm}=0$, for,
$m=p+1,\ldots,9$ and the configuration (3.19) will reduce to the
localized BPS D$p$-brane solutions.

This therefore shows how the non-supersymmetric $p$-brane
solutions delocalized in two transverse spatial directions
eq.(3.1) can be interpreted as the interpolating solutions between
D$(p+2)$-${\bar{\rm D}}(p+2)$ systems (eqs.(3.11,3.12)), non-BPS
D$(p+1)$-branes delocalized in one direction (eqs.(3.15,3.16)) and
localized BPS D$p$-branes (eq.(3.19)). This picture is very
similar to the descent relations obtained by Sen for the tachyonic
kink and vortex solutions on the brane-antibrane systems
\cite{asone}. However, there are some differences, particularly,
for the intermediate state i.e. the non-BPS D$(p+1)$-branes
starting from the brane-antibrane systems. For the case of Sen's
descent relation the non-BPS D$(p+1)$-brane obtained as a kink
solution interpolating tachyon vacua is a localized one, whereas,
we obtain the non-BPS D$(p+1)$-branes delocalized in one direction
which, unlike the BPS branes, we do not know how to localize.

\sect{The delocalized solutions and T-duality}

In this section we will study the T-duality properties of the delocalized
solutions given in section 2. Since the solutions obtained in section 2, have
the interpretation of interpolating solutions of brane configurations whose
dimensionalities differ by one, similar to the T-duality transformations,
we study this property for comparison of the results obtained in section 3.
But before we study the T-duality of the non-supersymmetric $p$-brane
solutions delocalized in two transverse directions, we study the T-duality
for the solutions delocalized in one transverse direction obtained in
ref.\cite{luroyone}.

Let us write down here the non-supersymmetric $p$-brane solutions
delocalized in one transverse spatial direction in $d=10$, \bea
ds^2 &=& F^{\frac{p+1}{8}} (H{\tilde {H}})^{\frac{2}{6-p}}
\left(\frac{H}{\tilde H}\right)^{-\frac{2\d_2}{6-p}}\left(dr^2 +
r^2 d\Omega_{7-p}^2\right) + F^{-\frac{7-p}{8}}\left(-dt^2 +
\sum_{i=1}^p dx_i^2 \right)\nn & &
\qquad\qquad\qquad\qquad\qquad\qquad + F^{\frac{p+1}{8}}
\left(\frac{H}{\tilde H}\right)^{2\d_2} dx_{p+1}^2\nn e^{2\phi}
&=& F^{-\frac{p-3}{2}} \left(\frac{H}{\tilde
{H}}\right)^{2\delta_1},\qquad F_{[8-p]}\,\,\, =\,\,\, b\,\,{\rm
Vol}(\Omega_{7-p})\wedge dx_{p+1} \eea where the function $F(r)$
is as given in (2.6). One can localize the above solutions along
$x_{p+1}$ direction without taking T-duality when the parameters
satisfy certain condition. The resulting solution can be
identified with non-BPS D$(p+1)$-brane which is chargeless and was
studied in ref.\cite{luroyone}. Here we will localize the above
solutions by applying T-duality along $x_{p+1}$-direction. For
that purpose we first write the metric in (4.1) in the string
frame as, \bea ds_{\rm str.}^2 &=& e^{\phi/2} ds^2\nn &=&
F^{\frac{1}{2}} (H{\tilde {H}})^{\frac{2}{6-p}}
\left(\frac{H}{\tilde
H}\right)^{-\frac{2\d_2}{6-p}+\frac{\d_1}{2}} \left(dr^2 + r^2
d\Omega_{7-p}^2\right)
\nn & & +
F^{-\frac{1}{2}}\left(\frac{H}{\tilde H}\right)^{\frac{\d_1}{2}}
\left(-dt^2 + \sum_{i=1}^p dx_i^2 \right) + F^{\frac{1}{2}}
\left(\frac{H}{\tilde H}\right)^{\frac{\d_1}{2}+2\d_2} dx_{p+1}^2
\eea After making a T-duality transformation \cite{bmm,bho,dr} the
$(p+1,p+1)$ component of the string frame metric in the dual
theory will be given as, \be {\tilde g}^{\rm str.}_{p+1,p+1} =
F^{-\frac{1}{2}}\left(\frac{H}{\tilde H}
\right)^{-\frac{\d_1}{2}-2\d_2} \ee The rest of the metric
components remain unaltered. The dilaton in the dual theory takes
the form \be e^{2\tilde \phi} =
F^{\frac{2-p}{2}}\left(\frac{H}{\tilde H}
\right)^{\frac{3\d_1}{2}-2\d_2} \ee We now rewrite the dual frame
metric from the string frame to the Einstein frame as, \bea
d{\tilde s}^2 &=& e^{-{\tilde\phi}/2} d{\tilde s}_{\rm str.}^2\nn
&=& F^{\frac{p+2}{8}} (H{\tilde {H}})^{\frac{2}{6-p}}
\left(\frac{H}{\tilde
H}\right)^{-\frac{2\d_2}{6-p}+\frac{\d_2}{2}+ \frac{\d_1}{8}}
\left(dr^2 + r^2 d\Omega_{7-p}^2\right)\nn & & +
F^{-\frac{6-p}{8}}\left(\frac{H}{\tilde H}\right)^{\frac{\d_2}{2}+
\frac{\d_1}{8}} \left(-dt^2 + \sum_{i=1}^p dx_i^2 \right) +
F^{-\frac{6-p}{8}} \left(\frac{H}{\tilde
H}\right)^{-\frac{7\d_1}{8}-\frac{3\d_2}{2}} dx_{p+1}^2 \eea Note
that in the above the dual frame metric $d{\tilde s}_{\rm str.}^2$
has the same form as in (4.2) with the $(p+1,p+1)$ component  as
given in (4.3)\footnote{From the magnetic charge as given in
(4.7), this configuration appears to represent the charged
D$(p+1)$-${\bar{\rm D}}(p+1)$ brane system. However, from the
following localization process, we see that this configuration
contains more than the above system. Examining the metric in (4.5)
and our experience in the brane bound state, we conclude that the above
configuration actually represents the charged D$(p+1)$-${\bar{\rm D}}(p+1)$
system with non-BPS
D$p$ brane uniformly distributed along $x_{p + 1}$ direction. In
other words, the above configuration represents a bound state of
charged (D$(p+1)$,${\bar{\rm D}}(p+1)$) system and non-BPS D$p$ branes. The
following localization process removes the delocalized non-BPS
D$p$ brane along the $x_{p + 1}$ direction.}. We observe that
(4.5) can indeed be made localized $(p+1)$-brane if we put\footnote{We
remark that all the $\delta$'s here need not be negative as otherwise
stated in our previous paper \cite{luroyone}. We noticed this after that
paper was published and this, however, does not change any of the conclusions
of the paper.} \be
\d_1 = -2 \d_2 \ee The complete solutions then take the forms,
\bea d{\tilde s}^2 &=& F^{\frac{p+2}{8}} (H{\tilde
{H}})^{\frac{2}{6-p}} \left(\frac{H}{\tilde
H}\right)^{\frac{\d_1}{6-p}-\frac{\d_1}{8} } \left(dr^2 + r^2
d\Omega_{7-p}^2\right) + F^{-\frac{6-p}{8}}\left(\frac{H}{\tilde
H}\right)^{-\frac{\d_1}{8} } \left(-dt^2 + \sum_{i=1}^{p+1} dx_i^2
\right)\nn e^{2\tilde \phi} &=&
F^{\frac{2-p}{2}}\left(\frac{H}{\tilde H}
\right)^{\frac{5\d_1}{2}}, \qquad F_{[7-p]}\,\,\,=\,\,\, b {\rm
Vol} (\Omega_{7-p}) \eea The parameters are related as, \bea
& &\a-\b\,\,\, =\,\,\, a\d_1\\
& & b\,\,\, =\,\,\, (6-p) (\a+\b) \omega^{6-p} \sinh2\theta\\
& & \frac{1}{2}\d_1^2 + \frac{\d_1^2}{4} \frac{7-p}{6-p} +
\frac{\a(\a - a\d_1)}{2}\,\,\,=\,\,\, \frac{7-p}{6-p} \eea So,
there are only three independent parameters characterizing the
solutions namely, $\d_1$, $\omega$ and $\theta$. We point out the
unlike the localization obtained in ref.\cite{luroyone}, the
localization obtained here by T-duality give charged $(p+1)$-brane
solutions. Also if the original delocalized $p$-brane solutions
belong to type IIA (IIB) theory, then the localized solutions
(4.7) obtained by T-duality belong to type IIB (IIA) theory. We
identify the solutions (4.7) in the dual theory as the charged
D$(p+1)$-${\bar{\rm D}}(p+1)$ brane systems \cite{luroytwo}. Let
us recall that the chargeless solutions obtained in
\cite{luroyone} without T-duality were identified as non-BPS
D$(p+1)$-branes. Comparing the solutions (4.7) with the
non-supersymmetric, charged $(p+1)$-brane solutions obtained in
eq.(2.30) of ref.\cite{luroytwo} in $d=10$, we find that they
match exactly if we identify, \bea {\hat F} &=& F
\left(\frac{H}{\tilde H}\right)^{\frac{\d_1}{6-p}}\nn {\hat \d}
&=& \frac{7-p}{6-p} \d_1\nn {\hat \alpha} &=& \alpha +
\frac{\d_1}{6-p} \eea where we have denoted the function $F$ and
the parameters in the solutions of ref.\cite{luroytwo} with a
`hat' to avoid any confusion. We can easily check that the
parameter relation (4.10) indeed match with the relation (2.26) of
ref.\cite{luroytwo} in $d=10$ under the above identifications. We
have thus shown that starting from the delocalized (in one
transverse direction) non-supersymmetric D$p$-brane solutions we
can obtain a localized, charged D$(p+1)$-${\bar{\rm
D}}(p+1)$-brane systems provided the parameters in the original
solutions satisfy eqs.(4.6). Conversely, one can also start from
charged D$(p+1)$-${\bar{\rm D}}(p+1)$-brane systems and applying
T-duality along $x_{p+1}$-direction, obtain the delocalized
non-supersymmetric D$p$-branes given in ref.\cite{luroyone} if the
parameters are identified as in (4.11) and satisfy (4.6).

Having described the T-duality properties of the
non-supersymmetric D$p$ branes delocalized in one transverse
direction, we proceed to study the T-duality of the delocalized
solutions in two transverse directions given in (3.1). In the
previous section we saw how the delocalized solutions can be made
localized without taking T-duality and the resulting solutions
were identified as D$(p+2)$-${\bar{\rm D}}(p+2)$-brane systems
with zero net charge. However, we know that the same theory also
contains D$(p+2)$-${\bar{\rm D}}(p+2)$-brane solutions with
non-zero RR charges. In this section we will show that the
solutions (3.1) can also be localized by taking T-duality twice
along the delocalized directions $x_{p+1}$ and $x_{p+2}$ and this
procedure will produce the charged D$(p+2)$-${\bar{\rm
D}}(p+2)$-brane systems that we just mentioned. In order to apply
T-duality let us rewrite the solutions (3.1) in the string frame
as, \bea ds_{\rm str.}^2 &=& e^{\phi/2} ds^2\nn &=&
F^{\frac{1}{2}} (H{\tilde {H}})^{\frac{2}{5-p}}
\left(\frac{H}{\tilde
H}\right)^{-\frac{2(\d_2+\d_3)}{5-p}+\frac{\d_1}{2}} \left(dr^2 +
r^2 d\Omega_{6-p}^2\right)\nn & & +
F^{-\frac{1}{2}}\left(\frac{H}{\tilde H}\right)^{\frac{\d_1}{2}}
\left(-dt^2 + \sum_{i=1}^p dx_i^2 \right) + F^{\frac{1}{2}}
\left(\frac{H}{\tilde H}\right)^{\frac{\d_1}{2}+2\d_2} dx_{p+1}^2
+ F^{\frac{1}{2}} \left(\frac{H}{\tilde
H}\right)^{\frac{\d_1}{2}+2\d_3} dx_{p+2}^2\nn e^{2\phi} &=&
F^{\frac{3-p}{2}}\left(\frac{H}{\tilde H}\right)^{2\d_1}, \qquad
F_{[8-p]} \,\,\,=\,\,\, b {\rm Vol}(\Omega_{6-p})\wedge dx_{p+1}
\wedge dx_{p+2} \eea Applying T-duality\footnote{
Although we do not explicitly write the configuration by taking T-duality
once along $x_{p+1}$ in (4.12), but, it can be easily shown that applying
T-duality in that case will lead to charged D$(p+1)$-${\bar{\rm D}}(p+1)$
brane system delocalized in $x_{p+2}$ direction and characterized by
four parameters.} \cite{bmm,bho,dr} along
$x_{p+2}$ and $x_{p+1}$, we obtain the $(p+2,p+2)$ and $(p+1,p+1)$
components of the string frame metric in the dual theory as, \bea
{\tilde g}^{\rm str.}_{p+2,p+2} &=&
F^{-\frac{1}{2}}\left(\frac{H}{\tilde H}
\right)^{-\frac{\d_1}{2}-2\d_3}\nn {\tilde g}^{\rm str.}_{p+1,p+1}
&=& F^{-\frac{1}{2}}\left(\frac{H}{\tilde H}
\right)^{-\frac{\d_1}{2}-2\d_2} \eea The rest of the metric
components remain the same. The dilaton in the dual theory takes
the form, \be e^{2\tilde \phi} =
F^{-\frac{p-1}{2}}\left(\frac{H}{\tilde H}
\right)^{\d_1-2\d_2-2\d_3} \ee Now we rewrite the T-dual solutions
in the same theory with the metric in the Einstein frame using
(4.12) -- (4.14) as\footnote{By the same token as given in the
footnote 6, the following configuration actually represents
a rather complicated bound state of charged (D$(p + 2)$, $\bar{\rm
D}(p + 2)$) system, non-BPS D$(p + 1)_1$, non-BPS D$(p + 1)_2$ and
chargeless (D$p$, $\bar{\rm D}p$) system where the subscripts `1' and
`2' refer to the non-BPS D$(p + 1)$ branes with one of their 
directions being along $x_{p + 1}$ and $x_{p +
2}$ respectively. The following localization process removes all
the branes but the the charged D$(p+2)$-${\bar{\rm D}}(p+2)$ brane
system.}, \bea d{\tilde s}^2 &=& e^{-{\tilde\phi}/2} d{\tilde
s}_{\rm str.}^2\nn &=& F^{\frac{p+3}{8}} (H{\tilde
{H}})^{\frac{2}{5-p}} \left(\frac{H}{\tilde
H}\right)^{-\frac{2(\d_2+\d_3)}{5-p}+\frac{\d_1}{4}+
\frac{(\d_2+\d_3)}{2}} \left(dr^2 + r^2 d\Omega_{6-p}^2\right)\nn
& &+ F^{-\frac{5-p}{8}}\left(\frac{H}{\tilde
H}\right)^{\frac{\d_1}{4}+ \frac{(\d_2+\d_3)}{2}} \left(-dt^2 +
\sum_{i=1}^p dx_i^2 \right) + F^{-\frac{5-p}{8}}
\left(\frac{H}{\tilde H}\right)^{-\frac{3\d_1}{4}-\frac{3\d_2}{2}
+ \frac{\d_3}{2}} dx_{p+1}^2\nn & & \qquad\qquad\qquad\qquad +
F^{-\frac{5-p}{8}} \left(\frac{H}{\tilde
H}\right)^{-\frac{3\d_1}{4}+\frac{\d_2}{2} - \frac{3\d_3}{2}}
dx_{p+2}^2\nn e^{2{\tilde \phi}} &=&
F^{-\frac{p-1}{2}}\left(\frac{H}{\tilde H}\right)^{
\d_1-2\d_2-2\d_3}, \qquad F_{[6-p]} \,\,\,=\,\,\, b {\rm
Vol}(\Omega_{6-p}) \eea Thus we note from above that the metric
can be localized if the coefficients of $(-dt^2 + \sum_{i=1}^p
dx_i^2)$, $dx_{p+1}^2$ and $dx_{p+2}^2$ match and this happens if
the parameters satisfy, \be \d_1 = -2\d_2 = -2\d_3 \ee So, the
localized solutions have only three independent parameters,
namely, $\omega$, $\theta$ and $\d_1$. The solutions then take the
forms, \bea d{\tilde s}^2 &=& F^{\frac{p+3}{8}} (H{\tilde
{H}})^{\frac{2}{5-p}} \left(\frac{H}{\tilde
H}\right)^{\frac{2\d_1}{5-p}-\frac{\d_1}{4} } \left(dr^2 + r^2
d\Omega_{6-p}^2\right) + F^{-\frac{5-p}{8}}\left(\frac{H}{\tilde
H}\right)^{-\frac{\d_1}{4} } \left(-dt^2 + \sum_{i=1}^{p+2} dx_i^2
\right)\nn e^{2\tilde \phi} &=&
F^{-\frac{p-1}{2}}\left(\frac{H}{\tilde H} \right)^{3\d_1}, \qquad
F_{[6-p]}\,\,\,=\,\,\, b {\rm Vol} (\Omega_{6-p}) \eea where the
parameters satisfy the relations, \bea & &
\a-\b\,\,\,=\,\,\,a\d_1\nn & & b \,\,\,=\,\,\, (5-p) \omega^{5-p}
(\a+\b) \sinh 2\theta\nn & & \frac{1}{2}\d_1^2 + \frac{1}{2}
\a(\a-a\d_1) + \frac{\d_1^2}{4(5-p)} \,\,\,=\,\,\,
\left(1-\frac{\d_1^2}{2}\right)\frac{6-p}{5-p} \eea We identify
the above solutions as the charged D$(p+2)$-${\bar {\rm
D}}(p+2)$-brane \cite{luroytwo} systems. Recall that the localized
solutions obtained in section 3 without taking T-duality were
chargeless and the corresponding solutions were identified as
chargeless D$(p+2)$-${\bar {\rm D}}(p+2)$-brane systems. Unlike
the case of solutions with one delocalized direction, the
localized solutions obtained here with or without T-duality belong
to the same theory i.e. either to type IIA (or IIB) theory because
we have taken T-duality twice. Indeed comparing the
non-supersymmetric, localized charged D$(p+2)$-${\bar {\rm
D}}(p+2)$-brane solutions given in eq.(2.30) of
ref.\cite{luroytwo} in $d=10$, we find that they match with (4.15)
provided, \bea {\hat F} &=& F \left(\frac{H}{\tilde
H}\right)^{\frac{2\d_1}{5-p}}\nn {\hat \d} &=&
\frac{2(7-p)}{(5-p)}\d_1\nn {\hat \a} &=& \a + \frac{2\d_1}{5-p}
\eea where we have denoted the quantities in the solutions (2.30)
of ref.\cite{luroytwo} with a `hat'. We can easily check that the
parameter relation given by the last equation of (4.18) goes over
to the parameter relation (2.26) of ref.\cite{luroytwo} in $d=10$
under the above identification. This concludes our discussion on
T-duality. The main difference between the localization obtained
without T-duality and with T-duality is that in the former case
the solutions are chargeless whereas, for the latter case the
solutions are charged.

\sect{Conclusion}

To summarize, we have obtained in this paper the
non-supersymmetric $p$-brane solutions delocalized in two
transverse spatial directions which arise as solutions of
$d$-dimensional supergravity containing a metric, a dilaton and a
$(d-p-3)$-form gauge field. The solutions are characterized
by five parameters. By adjusting and scaling the parameters we
have shown how these solutions in $d=10$ can be interpreted as
interpolating solutions between D$(p+2)$-${\bar{\rm
D}}(p+2)$-brane systems, non-BPS D$(p+1)$-branes (delocalized in
one direction) and localized BPS D$p$-branes. This picture is very
similar to the descent relations proposed by Sen  for the
tachyonic kink and vortex solutions on the brane-antibrane system.
For the case of descent relations all the brane configurations are
localized i.e. as the tachyon condenses, the energy of the system
can be shown to get localized to a codimension one brane at each
step. So, starting from a brane-antibrane system of dimensionality
$(p+2)$, one first gets a localized non-BPS brane of
dimensionality $(p+1)$ and then gets a localized BPS brane of
dimensionality $p$. However, in our case we do not get a localized
non-BPS D$(p+1)$-brane in the intermediate step. The reason is
that for the non-supersymmetric branes we do not know the relation
between the localized solution and the delocalized solution unlike
the case of BPS branes. The interpretation of the delocalized
solutions as the interpolating solutions and their similarity with
the tachyonic solitons obtained by Sen suggests that the dynamics
of tachyon condensation is perhaps related to the movements in the
parameter space associated with the solutions. However, to make
this statement more concrete it is necessary to give a microscopic
string interpretation of the solutions found in this paper and
relate the parameters with the physical parameters of the brane
system. For the case of localized solutions one such possible
interpretation was given in ref.\cite{luroythree,bmo} and it will
be interesting to find similar relations for the delocalized
solutions of this paper as well.

Since in interpreting the delocalized solutions as interpolating solutions
we have not used T-duality which also has the effect of localizing and
changing the dimensionality of the brane solutions by one, we have studied
this aspect in a separate section for comparison. We have studied the
T-duality of the solutions delocalized in both one and two transverse
directions. For the former case T-duality produces a localized
D$(p+1)$-${\bar{\rm D}}(p+1)$ brane system which is charged and changes the
theory from type IIA (IIB) to IIB (IIA). For the latter case T-duality
produces D$(p+2)$-${\bar{\rm D}}(p+2)$-brane system which is charged as opposed
to the chargeless solution one gets without T-duality. But in both cases
(with or without T-duality) the solutions belong to the same theory.

\vspace{.5cm}

\noindent {\bf Acknowledgements}

\vspace{2pt}

    JXL would like to thank the Michigan Center
for Theoretical Physics for hospitality during the final stage of
this work. He also acknowledges support by grants from the Chinese
Academy of Sciences and the grant from the NSF of China with Grant
No: 90303002.

\end{document}